\documentclass[12pt,onecolumn,prc,showpacs,preprintnumber,amsmath,
floatfix,amssymb]{revtex4}
\usepackage{epsfig}
\usepackage{epstopdf}
\usepackage{graphicx}
\usepackage{dcolumn}
\usepackage{bm}
\def\p{\mbox{\boldmath $p$}}
\def\q{\mbox{\boldmath $q$}}
\def\k{\mbox{\boldmath $k$}}

\allowdisplaybreaks[4]

\begin{document}
\title{Determination of the axial nucleon form factor from the MiniBooNE data}
\author{A.~V.~Butkevich$^{1}$ and D.~Perevalov$^{2}$}
\affiliation{ $^1$Institute for Nuclear Research,
Russian Academy of Sciences,
60th October Anniversary Prospect 7A,
Moscow 117312, Russia\\
$^2$Fermi National Accelerator Laboratory, Batavia, Illinois 60510, USA}
\date{\today}
\begin{abstract}

Both neutrino and antineutrino charged-current quasielastic scattering  
on a carbon target are studied to investigate the nuclear effect on the 
determination of the axial form factor $F_A(Q^2)$. A method for extraction of 
 $F_A(Q^2)$ from the flux-integrated $d\sigma/dQ^2$ cross section of 
(anti)neutrino scattering on nuclei is presented. Data from the MiniBooNE 
experiment are analyzed in the relativistic distorted-wave impulse 
approximation, the Fermi gas model, and the Fermi gas model with enhancements 
in the transverse cross section. We found that the values of the axial form 
factor, extracted in the impulse approximation and predicted by the dipole 
approximation with the axial mass $M_A\approx 1.37$, GeV are in good agreement. 
The agreement between the extracted form factor and meson-dominance ansatz also
 is good.
On the other hand, the $Q^2$ dependence of $F_A$ extracted in the approach 
with the transverse enhancement is found to differ significantly from the 
dipole approximation.
\end{abstract}

\maketitle

\section{Introduction}

One of the main systematic errors in the neutrino experiments are those 
associated with neutrino cross sections. 
The high-intensity neutrino beams used in neutrino oscillation experiments are 
peaked in the 0.3-5 GeV energy domain and allow the study of neutrino-nucleus 
interaction with unprecedented detail. In this energy 
range the dominant contribution to neutrino-nucleus scattering comes from the 
charged-current quasielastic (CCQE) reactions and resonance production 
processes.  

Various modern~\cite{NOMAD, MiniB1, MiniB2, MiniB3, Miner1, Miner2, T2K}
neutrino experiments reported measurements of the differential 
$d\sigma/d Q^2$ ($Q^2$ is squared four-momentum transfer)~\cite{NOMAD, MiniB1, 
MiniB2, MiniB3, Miner1, Miner2}, 
double-differential~\cite{MiniB1, MiniB3, T2K} and total cross 
sections~\cite{NOMAD, MiniB1, MiniB3, SciB, T2K} of CCQE scattering. This 
interaction represents a two-particle scattering process with lepton and 
nucleon in the final state. Since the criteria used to 
select CCQE events are strongly influenced by both target material and  
detector technology, various selection techniques are applied in these 
experiments (tracking and Cerenkov detector). In neutrino scattering 
experiments, the neutrino energy is unknown. However, both the neutrino energy
 and $Q^2$ can be evaluated based on the kinematics of the outgoing final state 
particles. 

With the assumptions of conserved vector current and partially conserved axial 
current, the only undetermined form factor in the theoretical description of 
the CCQE neutrino scattering is the axial nucleon form factor. In most analysis 
of neutrino interaction on complex nuclei, the dipole parametrization of 
$F_A(Q^2)$ with one parameter, the axial mass $M_A$, is used. The analyses are 
mainly based on the relativistic Fermi gas model (RFGM)~\cite{Smith} and 
involve some additional model dependence due to nuclear structure. While 
constraints exist from pion electroproduction data with $M_A=1.06\pm 0.016$ GeV~
\cite{Bernard}, neutrino experiments usually treat axial mass $M_A$ in CCQE as 
 independent of that measurement. 

The values of $M_A$ are obtained from a fit to observed $Q^2$ distribution of 
events, differential and total (anti)neutrino CCQE interaction cross sections. 
The formal averaging of $M_A$ values which are very widely spread was done in 
Ref.~\cite{Bernard}: $M_A=1.026\pm 0.021$ GeV. This result is also known as 
world-averaged value of the axial mass. Values of $M_A$ determined from the 
recent QE (anti)neutrino-carbon scattering experiments range anywhere from 
$M_A\approx$ 1 to 1.35 GeV. The NOMAD Collaboration~\cite{NOMAD} reports 
$M_A=1.05\pm 0.02\pm 0.06$ GeV, and the most resent MINERvA~\cite{Miner1, 
Miner2} results are consistent with $M_A \simeq 0.99$ GeV. On the other hand, 
the MiniBooNE Collaboration~\cite{MiniB1, MiniB2, MiniB3} reports a large value 
of $M_A=1.35\pm 0.17$ GeV. The other recent results~\cite{K2KSciFi, K2KSciBar, 
MINOS1, MINOS2} similarly find central values higher than the above-mentioned 
world average. The absolute values of the differential and total cross sections
 measured by MiniBooNE are about 30\% larger as compared to the NOMAD 
results. It is essential to obtain consistency between experiments utilizing 
different beam energies, nuclear targets, and detectors.

The radius of the nucleon axial charge distribution in terms 
of the dipole mass is defined as $\langle r^2_A\rangle=12/M^2_A$, and for 
$1.032 \le M_A \le 1.35$ GeV it can be estimated as $0.51 \le r_A \le 0.66$ fm.
So, a large value of $M_A=1.35$ GeV corresponds to a rather small axial charge 
radius of about 0.51 fm. On the other hand, the typical nucleon size as deduced 
from the electron scattering experiment is about 0.85 fm, whereas 
model-dependent analyses of proton-antiproton annihilation lead to a baryon 
charge radius of about 0.5 fm~\cite{Bernard}.

These results have encounraged many theoretical studies~\cite{Benh,Sob,BAV1,
BAV2,Meu3,Mart1,Mart2,Niev1,Niev2,Niev3,Amaro1,Amaro2,Amaro3,Bod1,
Lal} to attempt to explain the discrepancy between the data and 
traditional nuclear model, i.e., the RFGM. Some models that are based on the  
impulse approximation lead to 
large value of $M_A$~\cite{Benh,Sob,BAV1,BAV2} in the MiniBooNE data.
For instance, in Refs.~\cite{BAV1, BAV2}, a value of $M_A = 1.37$ GeV, that 
fits the $Q^2$ shape of the measured $d\sigma/d Q^2$ cross section was 
obtained within the RFGM, and relativistic distorted-wave impulse approximation 
(RDWIA) approaches. 

Other nuclear models~\cite{Mart1,Mart2,Niev1,Niev2,Niev3} include effects of 
multinucleon excitations such as meson-exchange currents (MECs) and isobar 
currents (ICs) to describe the MiniBooNE data. The contribution of the 
two-particle-two-hole (2p-2h) excitations to CCQE scattering has been found 
to be sizable and allows one to reproduce the MiniBooNE cross sections with 
value of 
$M_A\approx 1.03$ GeV. This result suggests that much of the cross section
 (about 30\%) measured by the MiniBooNE experiment can be attributed to 
processes that are not properly QE scattering. However, fully relativistic 
microscopic calculations of 2p-2h contributions are extremely difficult and 
may be bound to model-dependent assumptions. Future theoretical work is 
obviously needed to improve the present models that include effects beyond the 
impulse approximation. The transverse enhancement (TE) effective model to 
account for MEC effects has been proposed in Ref.~\cite{Bod1}. In this model, 
the magnetic form factors for nucleons bound in carbon are modified to describe
 the enhancement in the transverse electron-carbon QE cross section. The other 
nucleon form factors are the same as for free nucleons. It allows one to 
describe the MiniBooNE data (total cross section) by using the dipole 
approximation for $F_A(Q^2)$ with $M_A=1.014$ GeV.   

The assumption of the dipole ansatz for the axial form factor is a crucial 
element in these studies. The dipole parametrization of $F_A(Q^2)$ has no 
strict theoretical basis and the choice of this parametrization is made by 
analogy with electromagnetic form factors. On the other hand the dipole ansatz 
has been found to conflict with electron scattering data for the vector form 
factor~\cite{Bod1}. A model-independent description of the axial form factor 
was presented in Ref.~\cite{Hill}. The application of analyticity and 
dispersion relations to the axial form factor in the RFGM to find constraints 
for the axial mass using the data from MiniBooNE produces the value of 
$M_A\approx 0.85$ GeV which differs significantly from extractions based on the
traditional dipole ansatz. 

The pion and nucleon form factors were discussed in Ref.~\cite{Masjuan} within
the large-$N_c$ approach in the spacelike region. It was shown that, if the 
error bars on the monopole mass are taken into account one can make the dipole 
overlap with a product of monopoles within the corresponding error bars 
provided with the half-width rule. This construction is based on the following 
assumptions: (a) hadronic form factors in the spacelike region are dominated 
by mesonic states with the relevant quantum number, (b) the high-energy 
behavior is given by perturbative QCD, and the number of mesons is taken to be 
minimal to satisfy these conditions, and (c) errors in the meson-dominated form 
factors are estimated by means of the half-width rule.  

In Ref.~\cite{Bod2} the values of $F_A(Q^2)$ as a function of $Q^2$ were 
extracted from the differential cross section of neutrino scattering on 
deuterium (on a ``quasifree'' nucleon). A reasonable description of axial form 
factor by dipole approximation with $M_A=1.014\pm 0.016$ GeV was found. The 
current data on CCQE scattering come from a variety of experiments 
operating with carbon, oxygen, and iron data. Therefore the aim of this work is
 twofold. First, we propose a method which allow one to determine $F_A(Q^2)$ as
 a function of $Q^2$ directly from the measured flux-integrated 
$d\sigma /d Q^2$ cross section of CCQE (anti)neutrino scattering on nuclei. 
Second, we apply the method and extract the $Q^2$-dependence of the axial form 
factor from MiniBooNE data~\cite{MiniB1,MiniB3} in the RFGM, RDWIA and RFGM+TE 
approaches and show that the $Q^2$-dependence for the axial and vector form 
factors are correlated. We also compare the extracted $Q^2$-dependence of 
$F_A(Q^2)$ with predictions of the meson-dominance ansatz and dipole 
approximation predictions with $M_A\approx$ 1~GeV and $M_A\approx$ 1.3~GeV

The outline of this paper is the following. In Sec. II we present briefly the
RDWIA model and discuss the procedure which allows determination of the axial 
form factor from the flux-integrated $d\sigma/d Q^2$ (anti)neutrino cross 
section. Section III presents results of the extraction of $F_A$ from the 
MiniBooNE data and calculations of the flux-unfolded total cross sections for 
antineutrino scattering off carbon . Our conclusions are summarized in 
Sec. IV. 

\section{Model and method for extraction of $F_A(Q^2)$ from flux-integrated 
$d\sigma/dQ^2$ differential cross sections}

The formalism of CCQE exclusive

\begin{equation}\label{qe:excl}
\nu(k_i) + A(p_A)  \rightarrow \mu(k_f) + N(p_x) + B(p_B),      
\end{equation}
and inclusive
\begin{equation}\label{qe:incl}
\nu(k_i) + A(p_A)  \rightarrow \mu(k_f) + X                      
\end{equation}
scattering off nuclei in the one-W-boson exchange approximation has been 
extensively described in previous works~\cite{BAV3,BAV4,BAV5,BAV6}. Here 
$k_i=(\varepsilon_i,\k_i)$ and $k_f=(\varepsilon_f,\k_f)$ are the initial and 
final lepton momenta respectively, $p_A=(\varepsilon_A,\p_A)$, and 
$p_B=(\varepsilon_B,\p_B)$ are the initial and final target momenta, 
respectively $p_x=(\varepsilon_x,\p_x)$ is the ejectile nucleon momentum, 
$q=(\omega,\q)$ is the momentum transfer, and $Q^2=-q^2=\q^2-\omega^2$. 

\subsection{Model}

All the nuclear structure information and final 
state interaction (FSI) effects are contained in the weak CC nuclear tensors 
$W_{\mu \nu}$, which are given by a bilinear product of the transition matrix 
elements of the nuclear CC operator $J_{\mu}$ between the initial nucleus 
state and the final state.
We describe CCQE neutrino-nuclear scattering in the impulse approximation (IA),
 assuming that the incoming neutrino interacts with only one nucleon, which is 
subsequently emitted, while the remaining (A-1) nucleons in the target are 
spectators. The nuclear current is written as the sum of single-nucleon 
currents. 

The single-nucleon charged current has $V{-}A$ structure $J^{\mu} = 
J^{\mu}_V + J^{\mu}_A$. For a free-nucleon vertex function 
$\Gamma^{\mu} = \Gamma^{\mu}_V + \Gamma^{\mu}_A$ we use the vector 
current vertex function 
$\Gamma^{\mu}_V = F_V(Q^2)\gamma^{\mu} + 
{i}\sigma^{\mu \nu}q_{\nu}F_M(Q^2)/2m$, where 
$\sigma^{\mu \nu}=i[\gamma^{\mu}, \gamma^{\nu}]/2$ and $F_V$ and $F_M$ are the weak 
vector form factors. They are 
related to the corresponding electromagnetic ones for protons and neutrons by 
the hypothesis of the conserved vector current. We use the approximation 
of Ref.~\cite{MMD} on the nucleon form factors.
Because the bound nucleons are off shell we employ the de Forest 
prescription~\cite{deFor} and the Coulomb gauge for off-shell vector current 
vertex $\Gamma^{\mu}_V$. 

The axial current vertex function can be written in terms of the axial 
$F_A(Q^2)$ and the pseudoscalar $F_P(Q^2)$ form factors:
\begin{eqnarray}
\label{Eq3}
\Gamma_A^{\mu} &=& F_A(Q^2)\gamma^{\mu}\gamma_5 +F_P(Q^2)q^{\mu}\gamma_5   
\end{eqnarray}
The pseudoscalar form factor $F_P(Q^2)$ is dominated by the pion pole and is 
given in term of the Godberger-Treiman relation near $Q^2\approx 0$ if 
partially conserved axial current is assumed. We assume that the similar 
relation is valid for high $Q^2$ as well and write
\begin{eqnarray}
\label{Eq4}
F_p &=& \frac{2m^2F_A(Q^2)}{m^2_{\pi} +Q^2}=F_A(Q^2)F'_P(Q^2),   
\end{eqnarray}
where $F'_P(Q^2)=2m^2/(m^2_{\pi}+Q^2)$ and $m_{\pi}$ is the pion mass. Then the 
axial current vertex function can be written in the form
\begin{eqnarray}
\label{Eq5}
\Gamma_A^{\mu} &=& F_A(Q^2)[\gamma^{\mu}\gamma_5 +F'_P(Q^2)q^{\mu}\gamma_5].   
\end{eqnarray}
Thus the only undetermined form factor is the axial form factor that is 
commonly parametrized as a dipole 
\begin{eqnarray}
\label{Eq6}
F_A &=& \frac{F_A(0)}{(1 + Q^2/M^2_A)^2}   
\end{eqnarray}
where $F_A(0)$=1.267 and $M_A$ is the axial mass, which controls the $Q^2$ 
dependence of $F_A$, and ultimately, the normalization of the predicted cross 
section. 
We calculated the relativistic wave functions of the bound nucleon states  
in the independent particle shell model as the self-consistent solutions of a 
Dirac equation, derived within a relativistic mean field approach, from a 
Lagrangian containing $\sigma, \omega$, and $\rho$ mesons 
(the $\sigma-\omega$ model)\cite{Serot,Horow}. 
According to the JLab data~\cite{Dutta, Kelly1} the occupancy of the 
independent particle shell-model orbitals of ${}^{12}$C equals on 
average 89\%. In this work we assume that the missing strength (11\%) can be 
attributed to the short-range nucleon-nucleon ($NN$) correlations in the 
ground state, leading to the appearance of the high-momentum and 
high-energy component in the nucleon distribution in the target.  
These estimates of the depletion of hole states follow from the RDWIA 
analysis of ${}^{12}$C$({e},e^{\prime}{p})$ for 
$Q^2 < 2$ (GeV/c)$^2$~\cite{Kelly1} and are consistent with a direct 
measurement of the spectral function~\cite{Rohe}, which observed approximately 
0.6 protons in a region attributable to a single-nucleon knockout from a 
correlated cluster. 

In the RDWIA, final state interaction effects for the outgoing nucleon are 
taken into account. The distorted-wave function of the knocked out nucleon is 
evaluated as a solution of a Dirac equation containing a phenomenological 
relativistic optical potential. 
We use the LEA program~\cite{LEA} for the numerical calculation of the 
distorted wave functions with the EDAD1 parametrization~\cite{Cooper} of the 
relativistic optical potential for carbon. This code, initially designed for 
computing exclusive proton-nucleus and electron-nucleus scattering, was 
successfully tested against A$(e,e'p)$ data~\cite{Fissum, Dutta}, and we 
adopted this program for neutrino reactions.  

A complex optical potential with a nonzero imaginary part generally produces 
an absorption of the flux. For the exclusive A$(l,l'N)$ channel this reflects 
the coupling between different open reaction channels. 
However, for the inclusive reaction, the total flux must be conserved. 
Therefore, we calculate the inclusive and total cross sections with the EDAD1 
relativistic optical potential in which only the real part is included.

The inclusive cross sections with the FSI effects in the presence of the 
short-range $NN$ correlations were calculated by using the method proposed in
 Refs.~\cite{BAV3,BAV6}. In this approach the contribution of the $NN$ 
correlated pairs is evaluated in the IA, i.e., the virtual boson couples to only
 one member of the $NN$-pair. It is one-body process that leads to the emission 
of two nucleons (2p-2h excitation). The contributions of the 
two-body currents, such as meson-exchange currents and isobar currents,  are 
not considered. 

\subsection{Method for extraction of $F_A(Q^2)$ from flux-integrated 
$d\sigma/dQ^2$ cross sections}

The inclusive weak hadronic tensor $W_{\mu\nu}$ is given by bilinear products 
of the transition matrix elements of the nuclear weak current operators 
$J_{\mu}$ between the initial and final nuclear states, i.e., 
$W_{\mu\nu} = \langle J_{\mu} J^{\dagger}_{\nu}\rangle $, where the angle 
brackets denote products of matrix elements, appropriately averaged over 
initial states and summed over final states.

By using Eq.~\eqref{Eq5}, the axial vector current can be factorized in the form
\begin{eqnarray}
\label{Eq7}
J_A &=& F_A(Q^2)J'_A(Q^2),   
\end{eqnarray}
where
\begin{eqnarray}
\label{Eq8}
J'_A &=& \gamma^{\mu}\gamma_5 +F'_P(Q^2)q^{\mu}\gamma_5,  
\end{eqnarray}
and the weak current can be expressed as $J = J_V +F_AJ'_A$. The expression for 
the hadronic tensor is then given by
\begin{eqnarray}
\label{Eq9}
W_{\mu\nu} &=& W^V_{\mu\nu} + F^2_A(Q^2)W^A_{\mu\nu} + hF_A(Q^2)W^{VA}_{\mu\nu},  
\end{eqnarray}
where $W^V_{\mu\nu}=\langle (J_V)_{\mu} (J_V)^{\dagger}_{\nu}\rangle$, 
$W^A_{\mu\nu}=\langle (J'_A)_{\mu} (J'_A)^{\dagger}_{\nu}\rangle$,  
$W^{VA}_{\mu\nu}=\langle (J_V)_{\mu} (J'_A)^{\dagger}_{\nu} + 
(J'_A)_{\mu} (J_V)^{\dagger}_{\nu}\rangle$, and $h$ is 1 for a neutrino and -1 for 
an antineutrino. 
Finally, contracting $W_{\mu\nu}$ with the lepton tensor we obtain the inclusive 
(anti)neutrino scattering cross section $d\sigma/dQ^2$ in terms of vector 
$\sigma^V$, axial $\sigma^A$, and vector-axial $\sigma^{VA}$ cross sections
\begin{eqnarray}
\label{Eq10}
\frac{d\sigma^{\nu,~\bar{\nu}}}{dQ^2}(Q^2,\varepsilon_i) &=& 
\sigma^V(Q^2,\varepsilon_i) + F^2_A(Q^2)\sigma^A(Q^2,\varepsilon_i) + 
hF_A(Q^2)\sigma^{VA}(Q^2,\varepsilon_i),
\end{eqnarray}
where $\sigma^V=d\sigma/dQ^2(F_A=0)$ and 
$\sigma^A=d\sigma/dQ^2(F_V=F_M=0,F_A=1)$.
The vector (axial) cross section is due to the vector (axial) component of the 
weak current and it can be calculated as the $d\sigma/dQ^2$ cross section with 
$F_A(Q^2)=0$ $(F_V(Q^2)=F_M(Q^2)=0, F_A(Q^2)=1)$. So, the cross section 
$\sigma^A(Q^2)$ does not depend on vector form factors, i.e., on the 
longitudinal or transverse QE response functions. The vector-axial cross 
section $\sigma^{VA}$, arising from the interference between the vector and 
axial currents can be written as
\begin{eqnarray}
\label{Eq11}
\sigma^{VA} &=& [\sigma(F_A=1) - \sigma^V - \sigma^A],  
\end{eqnarray}
where $\sigma(F_A=1)$ is the $d\sigma/dQ^2$ cross section, calculated with 
$F_A(Q^2)$=1.

In the simplest case of (anti)neutrino scattering off a free nucleon the cross 
sections $\sigma^V$, $\sigma^A$, and $\sigma^{VA}$ can be expressed in terms of 
the vector form factors $F_V$ and $F_M$. For instance~\cite{Llew},
\begin{eqnarray}
\label{Eq12}
\sigma^{VA} &=& \frac{G^2_F}{2\pi}\cos^2\theta_c\frac{Q^2}{m\varepsilon_i}
\left(1-\frac{Q^2}{4m\varepsilon_i}\right)[F_V(Q^2)+F_M(Q^2)],  
\end{eqnarray}
where $G_F$ is the Fermi constant and $\theta_c$ is the Cabibbo angle. The
difference 
\begin{eqnarray}
\label{Eq13}
\frac{d\sigma^{\nu}}{dQ^2}(Q^2,\varepsilon_i)-
\frac{d\sigma^{\bar{\nu}}}{dQ^2}(Q^2,\varepsilon_i) &=& 
2F_A(Q^2)\sigma^{VA}(Q^2,\varepsilon_i),
\end{eqnarray}
is going to zero at $Q^2 \to$0 and decreases with (anti)neutrino energy. 

Our RDWIA results for the $\sigma^V$, $\sigma^A$, and $\sigma^{VA}$ cross 
sections for neutrino CCQE scattering off carbon are shown in Fig. 1 as 
functions of $Q^2$ for neutrino energies $\varepsilon_{\nu}$=0.5, 0.7, 1.2, and 
2.5 GeV. The cross section $\sigma^V$ has a maximum at 
$Q^2\approx 0.15$ (GeV/c)$^2$ and depends slowly on neutrino energy. The 
cross section $\sigma_A$ is dominant at $\varepsilon_{\nu} > 1$ GeV in the 
range $Q^2 > 0.2$ (GeV/c)$^2$
and slowly decreases with $Q^2$. On the other hand, the cross section 
$\sigma^{VA}$ decreases with neutrino energy as $\sim 1/\varepsilon_{\nu}$. 
At $\varepsilon_{\nu} > 1$ GeV, it depends slowly on $Q^2$ in the range 
$Q^2>0.3$ (GeV/c)$^2$.
\begin{figure*}
  \begin{center}
    \includegraphics[height=16cm,width=16cm]{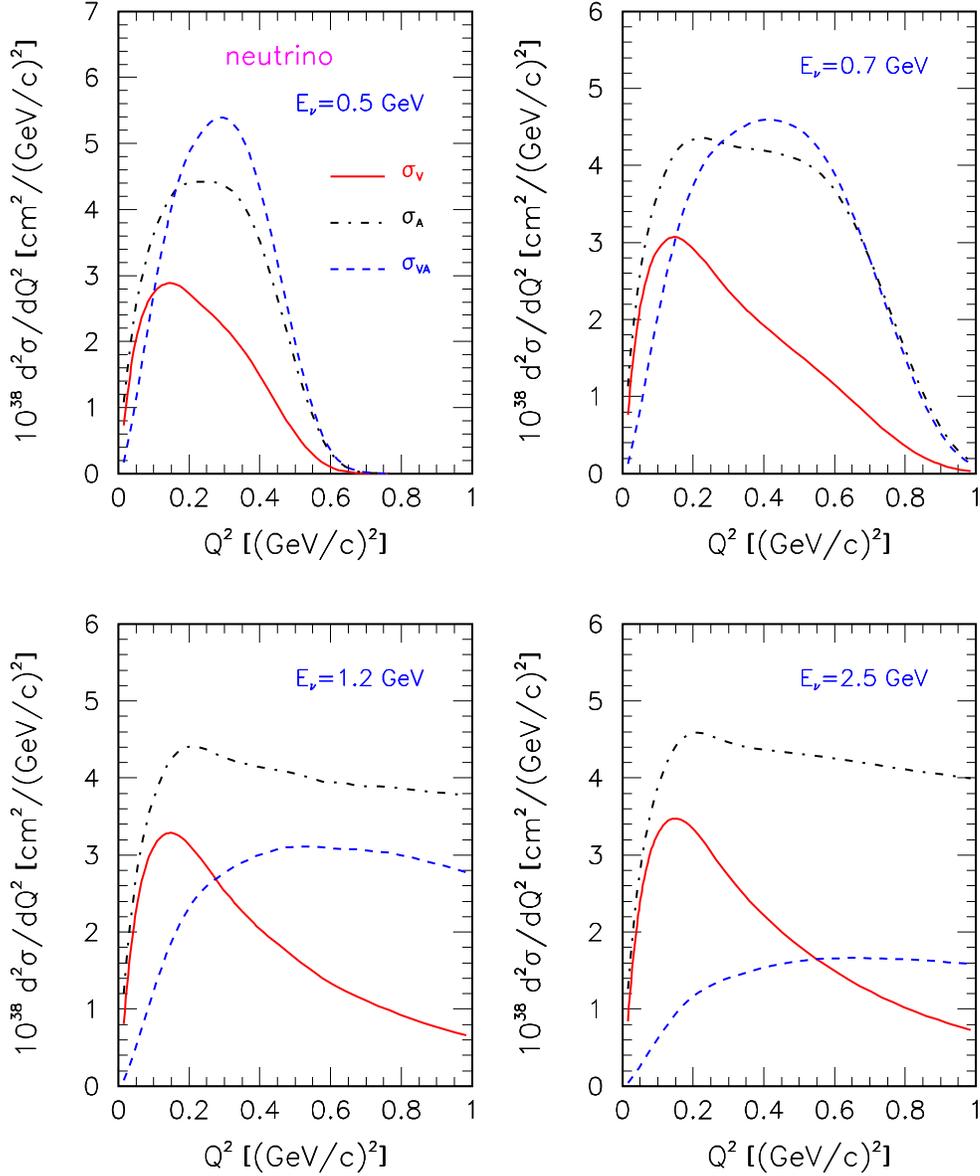}
  \end{center}
  \caption{\label{Fig1}(Color online) Differential cross sections $\sigma_V$ 
(solid line), $\sigma_A$ (dashed-dotted line), and $\sigma_{VA}$ (dashed line) 
vs four-momentum transfer $Q^2$ for neutrino scattering off carbon calculated 
in the RDWIA approach for four values of incoming neutrino energy: 
$\varepsilon_{\nu}=0.5, 0.7, 1.2$ and 2.5 GeV.} 
\end{figure*}
 
In neutrino experiments the differential cross sections of CCQE 
neutrino-nucleus scattering are measured within rather wide ranges of the 
(anti)neutrino energy spectrum. Therefore, flux-averaged and flux-integrated 
differential cross sections can be extracted.
The MiniBooNE $\nu_{\mu}$ and $\bar{\nu}_{\mu}$ CCQE flux-integrated 
$\left\langle d\sigma^{\nu,\bar{\nu}}/dQ^2 \right\rangle$ cross sections 
were measured as functions of $Q^2$ in the range 
$0\leq Q^2 \leq 2$ (GeV/c)$^2$~\cite{MiniB1,MiniB3}. These cross sections can 
be written as 
\begin{eqnarray}
\label{Eq14}
\left\langle \frac{d\sigma^{\nu,\bar{\nu}}}{dQ^2}(Q^2)\right\rangle &=& 
\int_{\varepsilon_{min}}^{\varepsilon_{max}}W_{\nu,\bar{\nu}}(\varepsilon_i)
\frac{d\sigma^{\nu,\bar{\nu}}}{dQ^2}(Q^2,\varepsilon_i) 
d\varepsilon_i.
\end{eqnarray}
The weight functions $W_{\nu,\bar{\nu}}$ are defined as 
\begin{eqnarray}
\label{Eq15}
W_{\nu,\bar{\nu}}(\varepsilon_i) &=& I_{\nu,\bar{\nu}}(\varepsilon_i)/
\Phi_{\nu,\bar{\nu}},
\end{eqnarray}
where $I_{\nu,\bar{\nu}}(\varepsilon_i)$ is the neutrino (antineutrino) spectrum 
and $\Phi_{\nu,\bar{\nu}}$ is the neutrino (antineutrino) flux in 
$\nu(\bar{\nu})$-beam mode~\cite{Flux}, integrated over 
$0\leq \varepsilon_i \leq 3$ GeV. Combining Eq.~\eqref{Eq10} with 
Eq.~\eqref{Eq14} we obtain the flux-integrated $\langle 
d\sigma^{\nu,\bar{\nu}}/dQ^2\rangle$ cross section in terms of flux-integrated
 $\langle \sigma^V\rangle$, $\langle \sigma^A\rangle$, and 
$\langle \sigma^{VA}\rangle$ cross sections 
\begin{eqnarray}
\label{Eq16}
\left\langle\frac{d\sigma^{\nu,\bar{\nu}}}{dQ^2}(Q^2)\right\rangle &=& 
\langle\sigma^V(Q^2)\rangle^{\nu,\bar{\nu}} + 
F^2_A(Q^2)\langle\sigma^A(Q^2)\rangle^
{\nu,\bar{\nu}} + h F_A(Q^2)\langle\sigma^{VA}(Q^2)\rangle^{\nu,\bar{\nu}},
\end{eqnarray}
where
\begin{eqnarray}
\label{Eq17}
{\left\langle \sigma^j(Q^2)\right\rangle}^{\nu,\bar{\nu}} &=& 
\int_{\varepsilon_{min}}^{\varepsilon_{max}}W_{\nu,\bar{\nu}}(\varepsilon_i)
[\sigma^j(Q^2,\varepsilon_i)]^{\nu,\bar{\nu}} d\varepsilon_i
\end{eqnarray}
are the flux-integrated vector, axial, and vector-axial $(j=V,A,AV)$ cross 
sections. The values of $F_A(Q^2)$ can be extracted as the 
solution of Eq.~\eqref{Eq16}, by using the data for 
$\langle d\sigma^{\nu,\bar{\nu}}/dQ^2\rangle$. In the case of neutrino scattering 
on deuterium (quasifree nucleon) this procedure was applied in  
Ref.~\cite{Budd} for the extraction of $F_A(Q^2)$ as a function of $Q^2$.

Because the flux-integrated cross 
sections $\left\langle \sigma^j \right \rangle$ depend on the nuclear model, 
vector form factors, and the predicted (anti)neutrino flux, the extracted 
values of the axial form factor are model dependet and implicitly include the 
uncertainties in the $F_V$, $F_M$, and $(\bar{\nu}_{\mu})\nu_{\mu}$ flux.

\section{Results and analysis}

\subsection{CCQE flux-integrated $\langle d\sigma^i(Q^2)\rangle$ differential 
cross sections}

The MiniBooNE $\nu_{\mu}(\bar{\nu}_{\mu})$ CCQE flux-integrated differential 
cross sections $d\sigma/dQ^2$ were extracted as functions of $Q^2$ in the 
range $0 \leq Q^2\leq 2$ (GeV/c)$^2$~\cite{MiniB1,MiniB3}. The cross sections 
are scaled with the number of neutron (proton) in the target. 
\begin{figure*}
  \begin{center}
    \includegraphics[height=16cm,width=16cm]{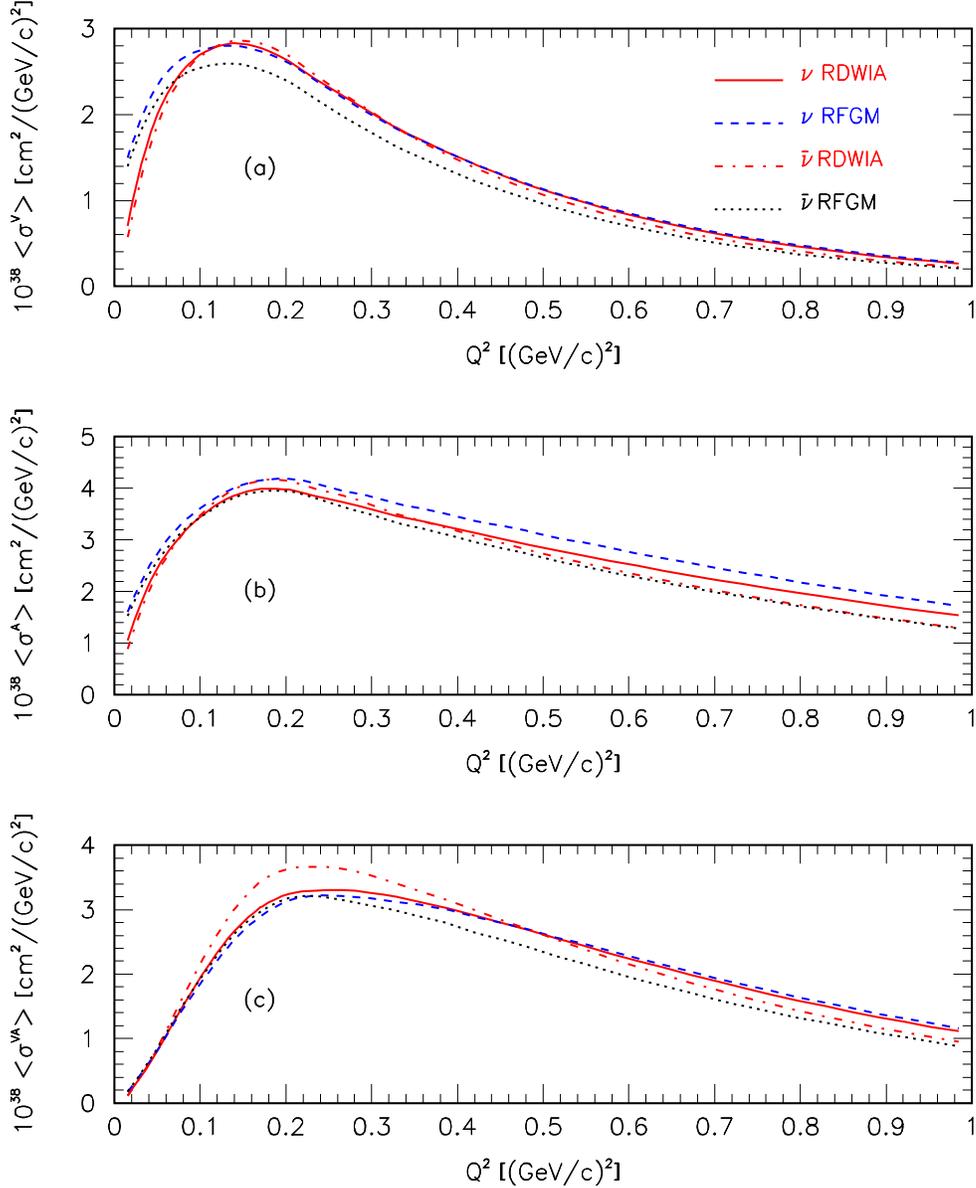}
  \end{center}
  \caption{\label{Fig2}(Color online) Flux-integrated 
$\langle \sigma^V\rangle^{\nu,\bar{\nu}}$, $\langle \sigma^A\rangle^{\nu,\bar{\nu}}$, 
and $\langle \sigma^{VA}\rangle^{\nu,\bar{\nu}}$  cross sections as functions of 
four-momentum transfer $Q^2$. As shown in the key, the cross sections were 
calculated in the RDWIA and RFGM with BNB $\nu_{\mu}$ and $\bar{\nu}_{\mu}$ 
fluxes.} 
\end{figure*}
A ``shape-only'' fit to the cross sections was performed
 to extract values for adjusted CCQE model parameters, $M_A$ and $\kappa$, 
within the Fermi gas model with the dipole parametrization of $F_A(Q^2)$. 
To tune this model to the low $Q^2$, the parameter $\kappa$~ was introduced.
The ``shape-only'' fit yields the model parameters, 
$M_A=1.35\pm 0.17$ (GeV/c)$^2$ 
and $\kappa=1.007\pm 0.012$. The extracted value for $M_A$ is approximately 
30\% higher than the world-averaged one. The prediction of the RFGM with these 
values of the parameters also describes well the measured (anti)neutrino 
flux-folded differential cross section $d\sigma^{\nu,\bar{\nu}}/dQ^2$.

To extract the values of the axial form factor $F_A(Q^2)$ as a function of 
$Q^2$ from the MiniBooNE neutrino and antineutrino flux-folded 
$d\sigma^{\nu,\bar{\nu}}/dQ^2$ cross sections we calculated the flux-integrated 
$\langle \sigma^V\rangle^{\nu,\bar\nu}$, $\langle \sigma^A\rangle^{\nu,\bar\nu}$, 
and $\langle \sigma^{VA}\rangle^{\nu,\bar\nu}$ cross sections with booster 
neutrino beam line (BNB) $\nu_{\mu}$ and $\bar{\nu}_{\mu}$ fluxes. In Fig. 2 
these cross sections, calculated within the RDWIA 
and RFGM (with the Fermi momentum $p_F=221$ MeV/c and a binding energy 
$\epsilon_b=25$ MeV for carbon), are shown as functions of $Q^2$. In the region
 $Q^2 \leq 0.25$(GeV/c)$^2$, the Fermi gas model results are higher than those 
obtained within the RDWIA. At $Q^2 \leq 0.05$(GeV/c)$^2$ this discrepancy 
equals $\sim 10$\% for $\langle \sigma^V\rangle$, $\sim 12$\% for 
$\langle \sigma^A\rangle$, and $\sim 8$\% for $\langle \sigma^{VA}\rangle$. 
To extract values for the axial form factor we calculated 
$\langle \sigma^i \rangle$ $(i=V,A,VA)$ cross sections with the BNB flux using 
the $Q^2$ bins $\Delta Q^2=Q^2_{i+1}-Q^2_i$ similar to Refs.~\cite{MiniB1,MiniB2}
\begin{eqnarray}
\label{Eq18}
\langle\sigma^i\rangle_j=
\frac{1}{\Delta Q^2}\int_{Q^2_j}^{Q^2_{j+1}}
\langle\sigma^i(Q^2)\rangle dQ^2 
\end{eqnarray}

\subsection{Extraction of the axial form factor}

Flux-integrated $\langle d\sigma^{\nu}/dQ^2\rangle$ and 
$\langle d\sigma^{\bar{\nu}}/dQ^2\rangle$ cross sections for 
$\nu_{\mu}$ and $\bar{\nu}_{\mu}$ CCQE scattering as functions of $Q^2$ together 
with the MiniBooNE data~\cite{MiniB1,MiniB3} are shown in Figs. 3 and 4 
(upper panel) correspondingly. The neutrino (antineutrino) cross section are 
scaled with the number of neutrons (protons) in the target.
\begin{figure*}
  \begin{center}
    \includegraphics[height=16cm,width=16cm]{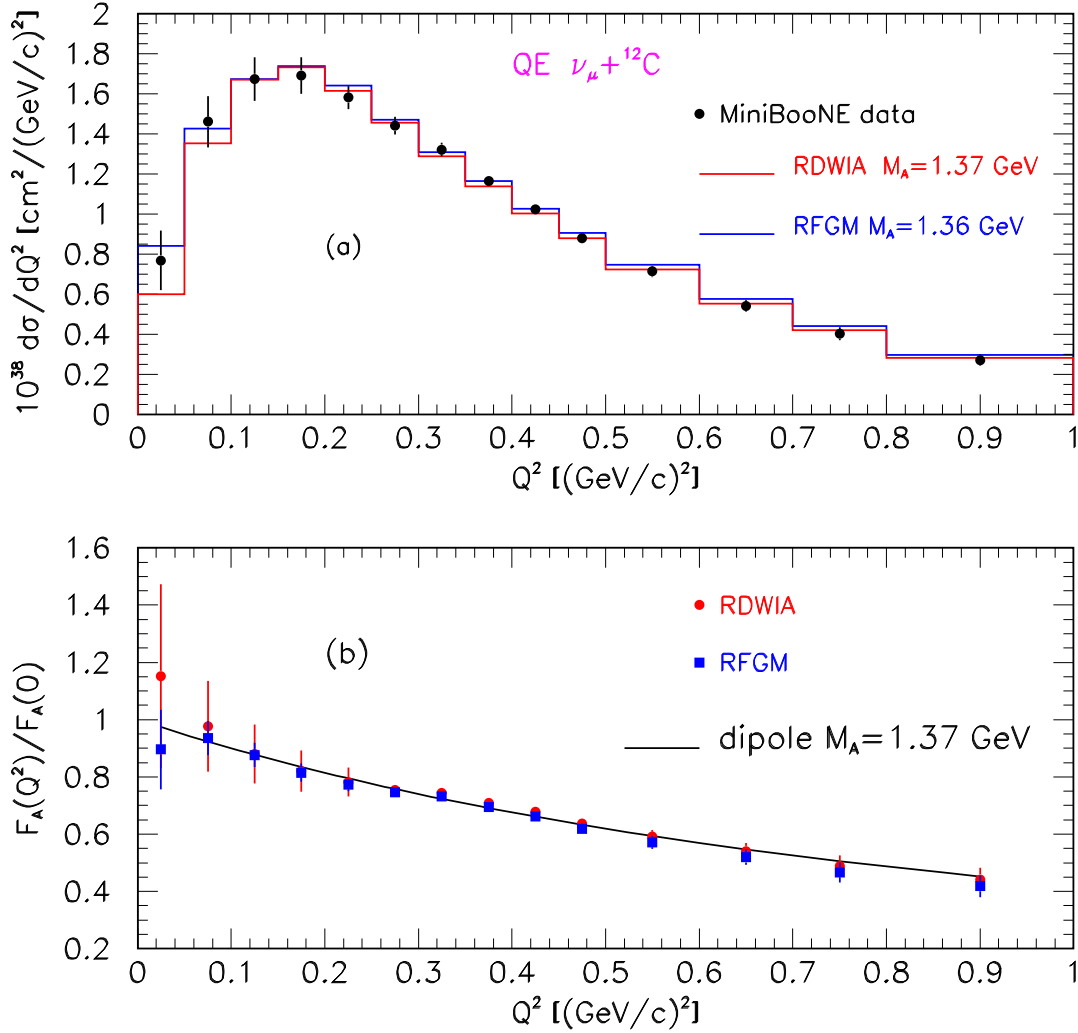}
  \end{center}
  \caption{\label{Fig3}(Color online) Flux-integrated 
$\langle d\sigma/dQ^2\rangle^{\nu}$ cross 
section per neutron target for the $\nu_{\mu}$ CCQE scattering (upper panel) and
 the normalized axial form factor $F_A(Q^2)/F_A(0)$ extracted from the 
MiniBooNE data (lower panel). Upper panel: Calculations from the RDWIA with 
$M_A=1.37$ GeV and RFGM with $M_A=1.36$ GeV. Lower panel: Filled 
circles (filled squares) are the axial form factor extracted within the RDWIA 
(RFGM) and the solid line is the result of the dipole parametrization with 
$M_A=1.37$ GeV.} 
\end{figure*}
They are calculated in the RDWIA
 ($M_A=1.37$ GeV) and RFGM ($M_A=1.36$ GeV) approaches. Also shown 
in Figs. 3 and 4 are the normalized axial form factor $F_A(Q^2)/F_A(0)$ (lower 
panel) extracted within the RDWIA and Fermi gas model from the 
measured cross sections. 
\begin{figure*}
  \begin{center}
    \includegraphics[height=16cm,width=16cm]{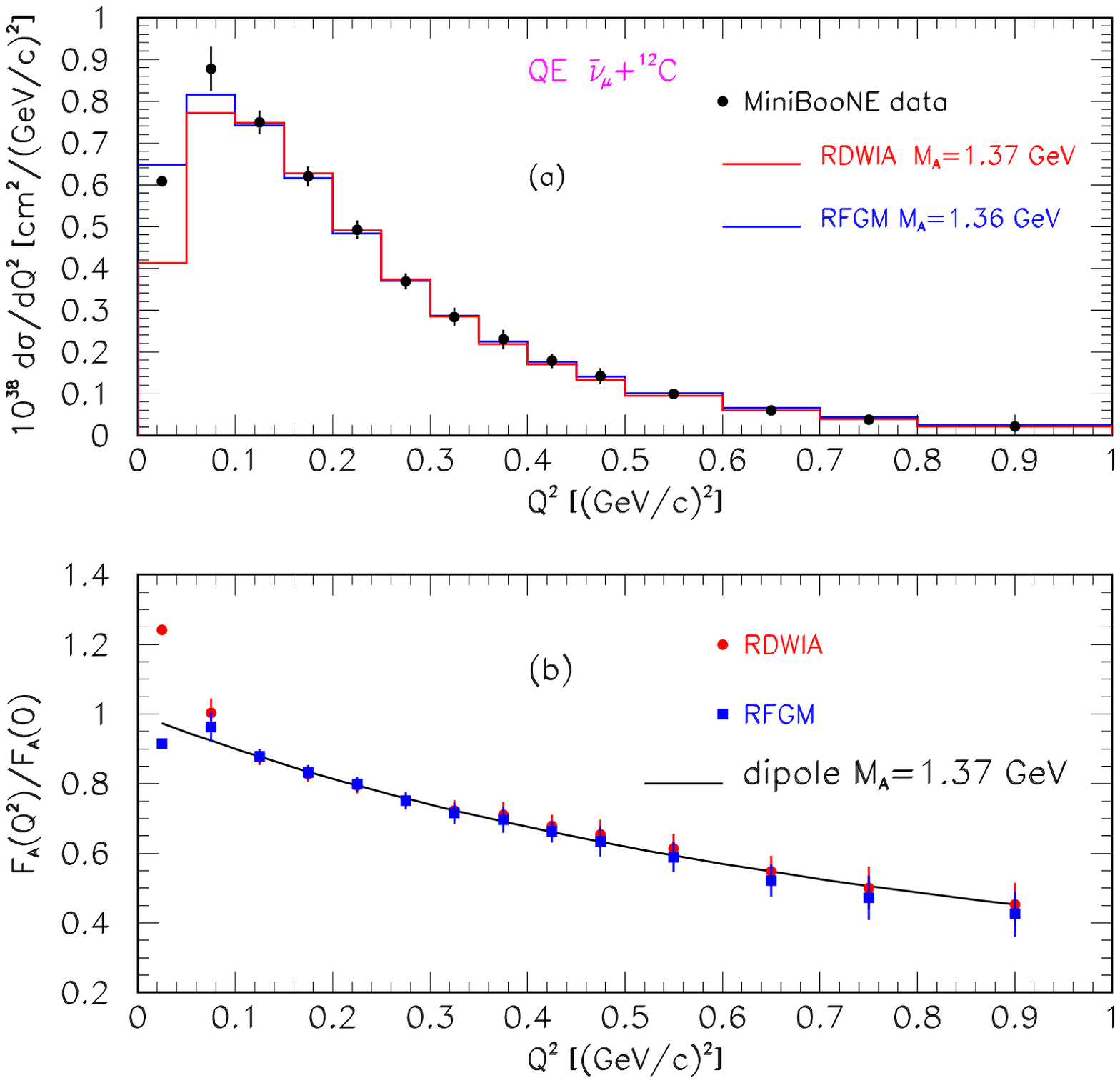}
  \end{center}
  \caption{\label{Fig4}(Color online) The same as Fig. 3, but for antineutrino 
scattering} 
\end{figure*}
The differences of the measured cross sections 
$\Delta [d\sigma/dQ^2]= \langle d\sigma/dQ^2\rangle^{\nu} - \langle d\sigma/dQ^2
\rangle^{\bar{\nu}}$ 
are shown in Fig. 5 (upper panel) as a function of $Q^2$ compared with the 
RDWIA and RFGM calculations. Also shown are the normalized axial form factors 
extracted in these approaches from $\Delta [d\sigma/dQ^2]$ (lower panel). The 
total normalization error on the neutrino (antineutrino) cross section 
measurement is 10.7\% (17.2\%). To extract values of $F_A(Q^2)$ the measured 
cross sections with `` the shape-only'' error were used in Eq.~\eqref{Eq16}.   
\begin{figure*}
  \begin{center}
    \includegraphics[height=16cm,width=16cm]{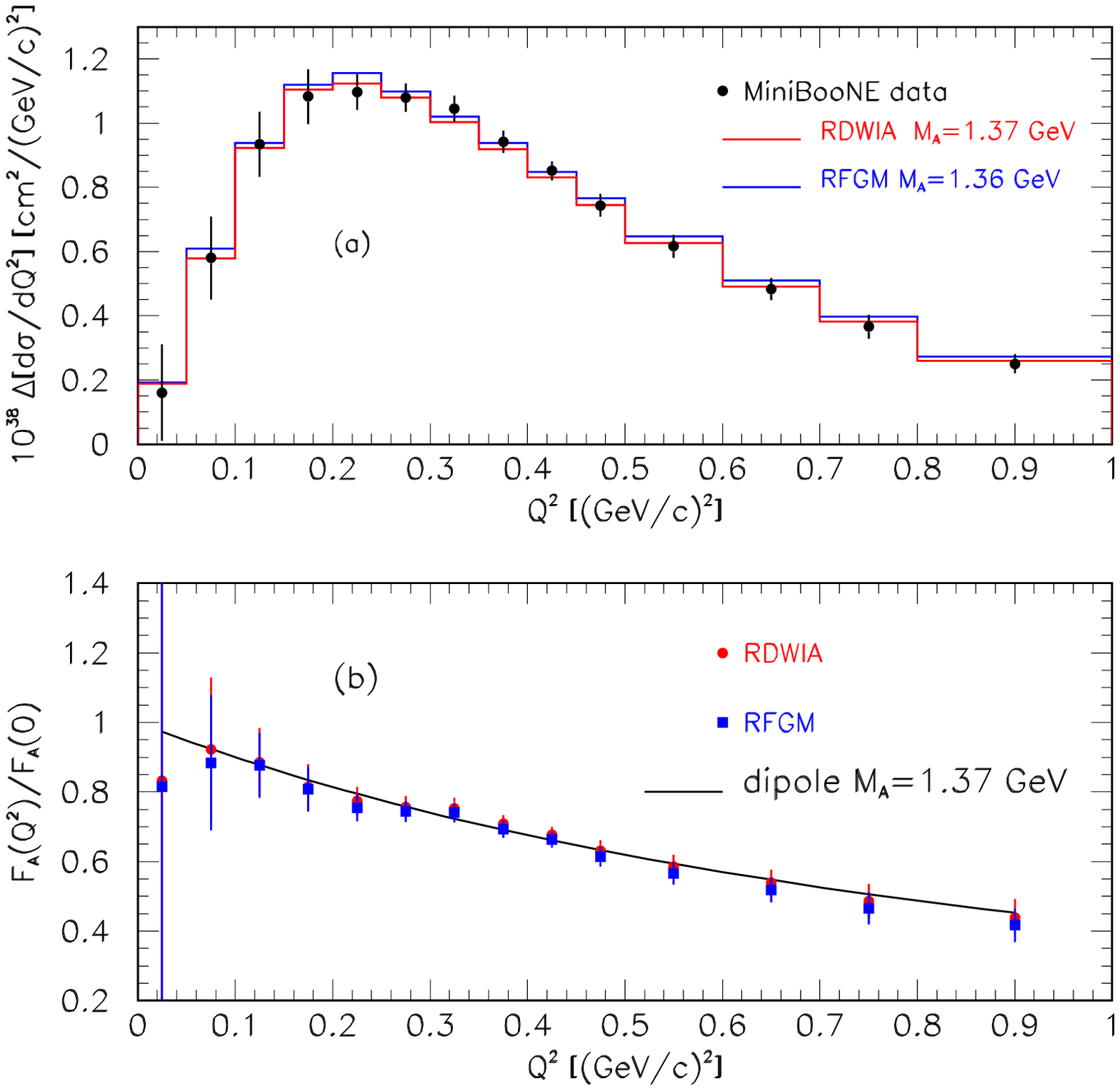}
  \end{center}
  \caption{\label{Fig5}(Color online) The same as Fig. 3 but for a difference 
of the measured cross sections $\Delta [d\sigma/dQ^2]$.} 
\end{figure*}

There is an overall agreement between the calculated and measured neutrino 
cross sections across the full range of $Q^2$. The values of the normalized 
axial form factors extracted within the RDWIA are similar to the RFGM result. 
A good match between the dipole parametrization with $M_A=1.37$ GeV and 
extracted form factors is observed. For antineutrino scattering there is an 
agreement between the calculated results and the data at $Q^2 \ge 0.1$ 
(GeV/c)$^2$ within the error of the experiment. However, the RDWIA result 
underestimates the measured cross section at $Q^2 < 0.1$ (GeV/c)$^2$. The 
values of $F_A(Q^2)$ extracted in the RDWIA and Fermi gas approaches at 
$Q^2 > 0.1$ (GeV/c)$^2$ also agree well with the dipole parametrization, 
whereas at $Q^2 \le 0.1$ (GeV/c)$^2$ the RDWIA result is higher than the RFGM 
and dipole parametrization prediction. A good match between the calculated 
and measured differences $\Delta [d\sigma/dQ^2]$ is observed. The extracted 
axial form factors also agree well with the dipole approximation 
prediction with $M_A=1.37$ GeV.

In Ref.~\cite{Bod1} it was assumed that enhancements in the transverse 
(anti)neutrino CCQE cross section are modified $F_M(Q^2)$, for a bound nucleon 
at low $Q^2\approx 0.3$ (GeV/c)$^2$.  The authors proposed a transverse 
enhancement function for the carbon target. If the TE   
originates from the MEC, then we may expect that enhancement in the 
longitudinal or axial contributions is small. Therefore in the TE model 
$F_V(Q^2)$ and $F_A(Q^2)$ are the same as for free nucleons. This approach was 
proposed to explain the apparent discrepancy between the low-(MiniBooNE) and 
high-(NOMAD) energy (anti)neutrino CCQE cross sections and $M_A$ measurements.

To study the TE effects on the extracted axial form factor we compare results 
of the RFGM ($M_A=1.36$ GeV) and the Fermi gas model ($M_A=1.014$ GeV) with the 
transverse enhancement function from Ref.~\cite{Bod1}. We call the last 
approach the RFGM + TE model~\cite{Sob2}. The flux-integrated cross sections 
$\langle d\sigma^{\nu}/dQ^2\rangle$ scaled with a number of neutron per target 
as functions of $Q^2$ together with the MiniBooNE data~\cite{MiniB1} are shown 
in Fig. 6. The upper panel shows cross sections calculated in the RFGM and 
RFGM+TE models. Also shown in Fig. 6 (lower panel) are normalized axial form 
factors extracted in these approaches from measured cross section.

In the range $Q^2\le 0.5$ (GeV/c)$^2$ the differential cross section  
$\langle d\sigma^{\nu}/dQ^2\rangle$ calculated within the RFGM+TE model is in 
agreement with the Fermi gas model prediction. However, in the region 
$Q^2\ge 0.7$ (GeV/c)$^2$ the RFGM+TE result is lower than the measured cross 
section. Therefore,
 there is a disagreement between the form factors extracted in this approach 
and predicted by the dipole parametrization with $M_A=1.014$ GeV. In the 
region $Q^2 \approx 0.2-0.3$ (GeV/c)$^2$ where an enhancement in the $F_M(Q^2)$ 
was assumed, the extracted values of $F_A$ are lower than those predicted from
 the dipole approximation. At $Q^2 > 0.6$ (GeV/c)$^2$ the enhancement function 
disappears with higher values of $Q^2$ and the value of $F_A$ extracted in the  
RFGM+TE model start approaching those extracted from the RFGM.
\begin{figure*}
  \begin{center}
    \includegraphics[height=16cm,width=16cm]{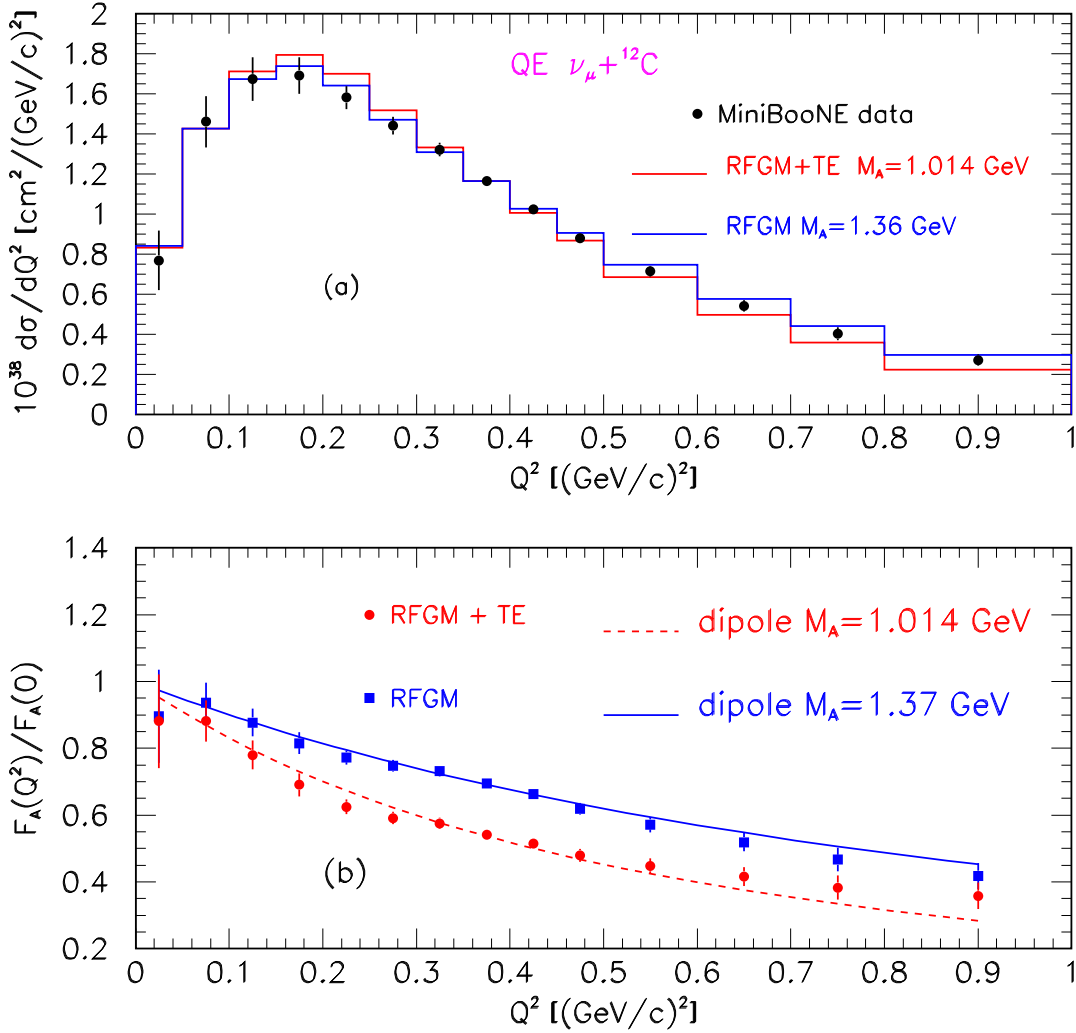}
  \end{center}
  \caption{\label{Fig6}(Color online) The same as Fig. 3, but calculated 
$\langle d\sigma^{\nu}/dQ^2\rangle$ cross sections per neutron target and 
extracted form factors are from the RFGM ($M_A=1.36$ GeV) and RFGM+TE 
($M_A=1.014$ GeV)} 
\end{figure*}
As shown in Fig. 6 the shape of the $Q^2$ dependence of the axial form factor 
extracted within the RFGM+TE approach cannot be well described by the dipole 
ansatz. So, assuming the dipole parametrization of $F_A(Q^2)$ the 
MiniBooNE data may be described within the impulse approximation with large 
axial mass value $\sim 1.3$ GeV as well as within the approaches that 
include sizable MEC and IC contributions and allow reproduction of data with 
$M_A\sim 1$ GeV. But for the self-consistent description of the MiniBooNE 
data with the enhancement in the transverse response function (at least as it 
was proposed in Ref.~\cite{Bod1}), one should use a parametrization of 
$F_A(Q^2)$ other than a dipole. 

As was pointed in Ref.~\cite{Masjuan}, the dipole approximation cannot be 
justified from a field-theoretic point of view and is in contradiction with 
the large-$N_c$-motivated parametrization. We calculated the normalized axial 
form factor by using the minimum meson-dominance ansatz from 
Ref.\cite{Masjuan}: 
\begin{eqnarray}
\label{Eq19}
F_A(Q^2)=F_A(0)\frac{m^2_{a_1}m^2_{a'_1}}{(m^2_{a_1} +Q^2)(m^2_{a'_1}+Q^2)}, 
\end{eqnarray}
where $m_{a_1}=1.230$ GeV, and $m_{a'_1}=1.647$ GeV. By applying the half-width 
rule $m_R \pm \Gamma_R/2$ to this parametrization with $\Gamma_{a_1}=0.425$ GeV 
and $\Gamma_{a'_1}=0.254$ GeV, we get results depicted in Fig. 7. The errors in 
the meson-dominanted form factor (the band for the form factor) are estimated 
by treating resonance masses as random variables distributed with the 
dispersion given by the width. As we can see, the agreement between the 
extracted form factor and the meson dominance ansatz is impressive. Actually, 
the two axial mesons are incorporated as a product of monopoles, but the net 
effect is essentially a dipole form factor with an average mass which is 
larger than $m_R=1$ GeV. 
\begin{figure*}
  \begin{center}
    \includegraphics[height=15cm,width=15cm]{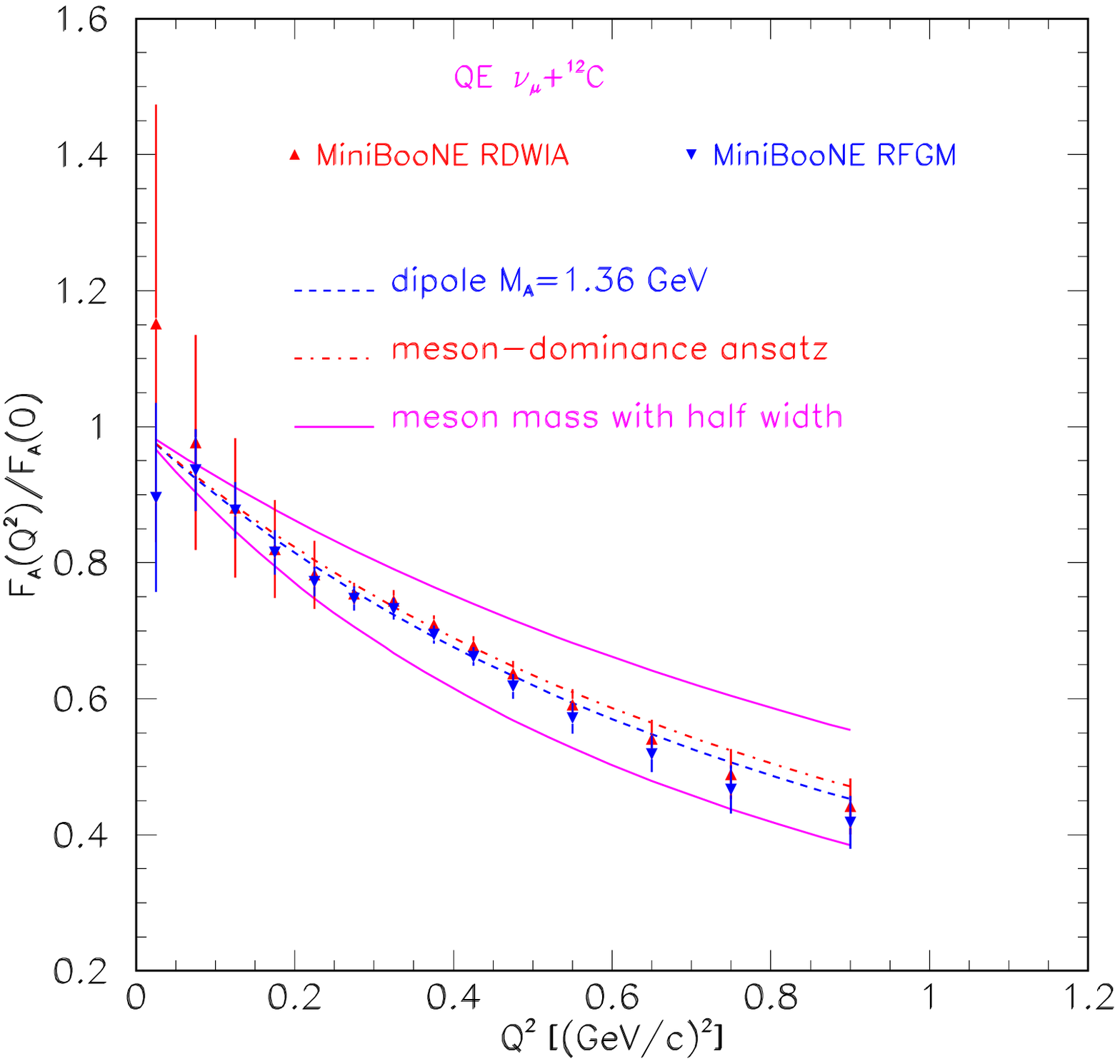}
  \end{center}
  \caption{\label{Fig7}(Color online) The normalized axial form factor 
$F_A(Q^2)/F_A(0)$ extracted from MiniBooNE data. The dashed line is the result 
of the dipole parametrization with $M_A=1.36$ GeV, dashed-dotted line is the 
result of the meson-dominance ansatz, and solid lines show the bands for the 
form factor due to the half-width of the meson masses.} 
\end{figure*}

\subsection{CCQE total cross section}

\begin{figure*}
  \begin{center}
    \includegraphics[height=17cm,width=17cm]{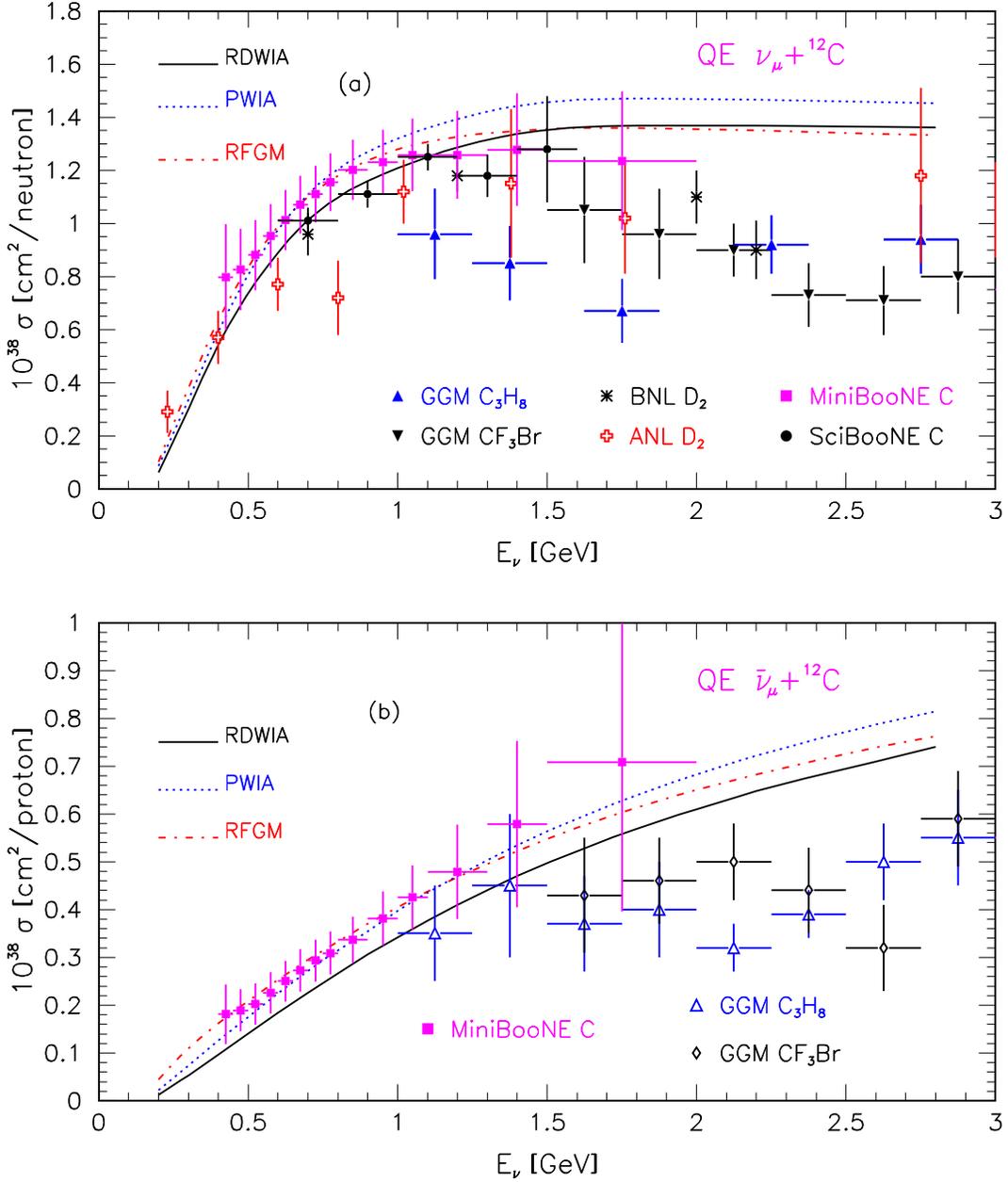}
  \end{center}
  \caption{\label{Fig8}(Color online) Total $\nu_{\mu} (\bar{\nu}_{\mu})$ CCQE 
cross section per neutron (proton) target as a function of neutrino energy. 
Data points for different targets are from Refs.~\cite{SciB,MiniB2,Mann,Baker,
Pohl,Brunner}. Also shown are predictions of the RDWIA ($M_A=1.37$ GeV), 
PWIA ($M_A=1.37$ GeV), and RFGM ($M_A=1.36$ GeV).}
\end{figure*}
We calculated the total cross sections for CCQE antineutrino scattering on 
carbon in the plane-wave impulse approximation (PWIA), RDWIA, and RFGM 
approaches. The cross section per proton target as a function of antineutrino 
energy is shown in Fig. 8 (lower panel) together with the data from  
Refs.~\cite{MiniB1,MiniB3,Mann, Baker, Pohl, Brunner}. Also shown are (upper 
panel) the total cross sections per neutron target for neutrino scattering 
from Ref.~\cite{BAV1}. The calculated cross sections, which use the values of 
$M_A$ extracted from the shape-only fit to the flux-integrated 
$d\sigma^{\nu,~\bar{\nu}}/Q^2$ data, reproduce the MiniBooNE total cross section 
within the errors. At the energy of $\varepsilon_{\nu}\approx 700$ MeV, the 
extracted cross section is $\approx 30\%$ higher than what is commonly assumed 
for this process assuming the RFGM and world-average value of the axial mass 
$M_A=1.03$ GeV. As shown in Fig. 7 the spread in the data is much higher than 
the difference in predictions of the RDWIA, PWIA, and RFGM approaches. 

\section{Conclusions}

In this paper, we present a method which allows extraction of the axial form 
factor as a function of $Q^2$ from flux-integrated 
$\langle d\sigma^{\nu,\bar{\nu}}/dQ^2\rangle$ cross sections. This method is based
 on the fact that the $d\sigma/dQ^2$ can be written as the sum of the vector, 
axial and vector-axial cross sections and the contributions of the axial and 
vector-axial ones are proportional to $F_A^2(Q^2)$ and $F_A(Q^2)$ 
correspondingly. In our analysis we used the differential 
$\langle d\sigma^{\nu,\bar{\nu}}/dQ^2\rangle$ CCQE cross sections with 
``a shape-only'' error measured in the MiniBooNE experiment. 

In the RDWIA, RFGM, and RFGM+TE approaches with the BNB $\nu_{\mu} 
(\bar{\nu}_{\mu})$ flux we calculated the flux-integrated vector, axial, and 
vector-axial cross sections which were used for extraction of the axial form 
factor. The values of $F_A(Q^2)$ extracted in the RDWIA and Fermi gas models at
 $Q^2 > 0.1$ (GeV/c)$^2$ agree well with the dipole parametrization with 
$M_A\approx 1.37$ GeV. 
On the other hand the vector meson-dominance ansatz that is simple and has good 
theoretical base describes the extracted form factor also with a good 
agreement. We can argue that there is no need to use the dipole approximation 
for fitting the MiniBooNE data and that the meson dominance already contains 
the essential physical information, whereas we found that there is a 
disagreement between the form factors extracted in the RFGM+TE approach and 
predicted by the dipole approximation with $M_A\approx 1.04$ GeV. The RDWIA, 
PWIA and RFGM calculated with $M_A=1.37$ GeV and measured neutrino and 
antineutrino CCQE total cross sections match well within the experimental 
error over the entire measured range. 

We conclude that the MiniBooNE measured inclusive and total cross sections can 
be described within the impulse approximation with the meson-dominance ansatz 
and/or dipole approximation of $F_A(Q^2)$ with large axial mass value.
In addition, one could also describe MiniBooNE data with the approach that 
incorporates the large MEC contributions. However, 
a non-dipole description of the axial from factor would have to be used in 
that case. From the quality of the fit to the measured MiniBooNE CCQE cross 
section, one could not discriminate between these approaches.

So, to distinguish between various possible mechanisms, it is necessary to 
decompose the inclusive process into its constituents exclusive channels, at 
least to study the semiexclusive $\nu_{\mu}+A \to \mu + p + B$ process. In this
 manner one hopes to disentangle the roles of $NN$ correlations in the target 
ground state, MEC, IC, medium modifications properties of the bound nucleon, 
many-body currents, FSI, relativistic corrections et cetera, 
etc.~\cite{Kelly2}

\section*{Acknowledgments}

The authors gratefully acknowledge P.~Masjuan, E.~R.~Arriola, W.~Broniowsky, 
J.~Amaro, and N.~Evans for fruitful discussion the results obtained 
within the meson dominance model and a critical reading of the manuscript.

%



\begin{thebibliography}{99}
\bibitem{NOMAD} V.~Lyubushkin {\it et al.}, (NOMAD Collaboration), 
Eur. Phys. J. {\bf C63}, 355, (2009).
\bibitem{MiniB1} A.~A.~Aguilar-Arevalo {\it et al.}, (MiniBooNE 
Collaboration), Phys. Rev. {\bf D81}, 092005 (2010).
\bibitem{MiniB2} A.~A.~Aguilar-Arevalo {\it et al.}, (MiniBooNE 
Collaboration), Phys. Rev. {\bf D82}, 092005 (2010).
\bibitem{MiniB3} A.~A.~Aguilar-Arevalo {\it et al.} (MiniBooNE Collaboration)
Phys. Rev. {\bf D88}, 032001 (2013).
\bibitem{Miner1} L.~Fields {\it et al.}, (MINERvA Collaboration), 
Phys. Rev. Lett. {\bf 111}, 022501 (2013).
\bibitem{Miner2} G.~A.~Fiorentini {\it et al.}, (MINERvA Collaboration), 
Phys. Rev. Lett. {\bf 111}, 022502 (2013).
\bibitem{T2K} K.~Abe {\it et al.}, (T2K Collaboration), Phys. Rev. {\bf D87}, 
092003 (2013).
\bibitem{SciB} Y.~Nakajima {\it et al.}, (SciBooNE Collaboration),
  Phys. Rev. {\bf D83}, 012005 (2011).
\bibitem{Smith} R.~A.~Smith and E.~J.~Moniz, Nucl. Phys.{\bf B43}, 605 (1972); 
\textit{erratum: ibid.} {\bf B101}, 547 (1975). 
\bibitem{Bernard} V.~Bernard, L.~E.~Elouadrhiri, U.~-G.~Meissner J. Phys. 
{\bf G28}, R1, (2002).
\bibitem{K2KSciFi} R.~Gran {\it et al.}, (K2K Collaboration), Phys. Rev. 
{\bf D74}, 052002 (2006). 
\bibitem{K2KSciBar} X.~Espinal, F.~Sanchez, AIP (Conf. Proc.) 
{\bf 967}, 117 (2007).
\bibitem{MINOS1} M.~Dorman {\it et al.} (MINOS Collaboration),  
  AIP Conf. Proc. {\bf 1189}, 133 (2009).
\bibitem{MINOS2} N.~S.~Mayer, PhD Thesis, Indiana Univ. (2011).
\bibitem{Benh} O.~Benhar, P.~Coletti, and D.~Meloni, Phys. Rev. Lett.{\bf 105},
 132301 (2010).
\bibitem{Sob} C.~Juszczak, J.~T.~Sobczyk, and J.~Zmuda, Phys. Rev. {\bf C82},
 045502 (2010).
\bibitem{BAV1} A.~V.~Butkevich, Phys. Rev. {\bf C82}, 055501 (2010).
\bibitem{BAV2} A.~V.~Butkevich, and D.~Perevalov, Phys. Rev. {\bf C84}, 015501 
(2011).
\bibitem{Meu3} A.~Meucci, C.~Giusti, Phys. Rev. {\bf D85}, 093002 (2012).
\bibitem{Mart1} M.~Martini, M.~Ericson, G.~Chanfray, and J.~Marteau, 
Phys. Rev. {\bf C81}, 045502 (2010)
\bibitem{Mart2} M.~Martini, M.~Ericson, and G.~Chanfray, Phys. Rev. {\bf C84}, 
055502 (2011)
\bibitem{Niev1} J.~Nieves, I.~R.~Simo and M.~J.~Vicente~Vacas, Phys. Rev. 
{\bf C83}, 045501 (2011).
\bibitem{Niev2} J.~Nieves, I.~Ruiz~Simo, and M.~J.~Vicente~Vacas, Phys. Lett. 
{\bf B707}, 72 (2012).
\bibitem{Niev3} J.~Nieves, I.~Ruiz~Simo, and M.~J.~Vicente~Vacas, Phys. Lett. 
{\bf B721}, 90 (2013).
\bibitem{Amaro1} J.~E.~Amaro, M.~B.~Barbaro, J.~A.~Caballero, T.~W.~Donnelly, 
and C.~F.~Williamson, Phys. Lett. {\bf B696}, 151 (2011).
\bibitem{Amaro2} J.~E.~Amaro, M.~B.~Barbaro, J.~A.~Caballero, T.~W.~Donnelly, 
J.~M.~Udias, Phys. Rev. {\bf D84}, 033004 (2011).
\bibitem{Amaro3} J.~E.~Amaro, M.~B.~Barbaro, J.~A.~Caballero, T.~W.~Donnelly, 
Phys. Rev. Lett. {\bf 108}, 152501 (2012).
\bibitem{Bod1} A.~Bodek, H.~Budd, and M.~Christy, Eur. Phys. J. {\bf C71}, 
1, (2011).
\bibitem{Lal} O.~Lalakulich, K.~Gallmeister, and U.~Mosel, Phys. Rev. 
{\bf C86}, 014614, (2012).
\bibitem{Hill} B.~Bhattacharya, R.~J.~Hill, and G.~Paz, Phys. Rev. {\bf D84}, 
073006 (2011).
\bibitem{Masjuan} P.~Masjuan, E.~R.~Arriola, and W.~Broniowski, Phys. Rev. 
{\bf D87}, 014005 (2013).
\bibitem{Bod2} A.~Bodek, S.~Avvakumov, R.~Bradford, and H.~Budd, Eur. Phys. J. 
{\bf C53}, 349, (2008).
\bibitem{BAV3} A.~V.~Butkevich and S.~A.~Kulagin, Phys. Rev. {\bf C76}, 045502
 (2007).
\bibitem{BAV4} A.~V.~Butkevich, Phys. Rev. {\bf C78}, 015501 (2008).
\bibitem{BAV5} A.~V.~Butkevich, Phys. Rev. {\bf C80}, 014610 (2009).
\bibitem{BAV6} A.~V.~Butkevich, Phys. Rev. {\bf C85}, 065501 (2012).
\bibitem{MMD} P.~Mergell, U.-G.~Meissner, and D.~Drechesel, Nucl. Phys.
{\bf A596}, 367, 1996.
\bibitem{deFor} T.~de~Forest, Nucl. Phys. {\bf A392}, 232, 1983.
\bibitem{Serot} B.~Serot, J.~Walecka, Adv. Nucl. Phys. {\bf 16}, 1, 1986.
\bibitem{Horow} C.~J.~Horowitz D.~P.~Murdock, and Brian~D.~Serot, in 
{\it Computational Nuclear Physics 1: Nuclear Structure} edited 
by K.~Langanke, J.~A.~Maruhn, Steven~E.~Koonin (Springer-Verlag,Berlin, 1991), 
p.129
\bibitem{Dutta} D.~Dutta {\it et al.}, Phys. Rev. {\bf C68}, 064603, (2003).
\bibitem{Kelly1} J.~J.~Kelly Phys. Rev. {\bf C71}, 064610 (2005).
\bibitem{Rohe} D.~Rohe {\it et al.}, Nucl. Phys. B (Proc. Suppl.) 
{\bf 159}, 152 (2006).
\bibitem{LEA} J.~J~Kelly, http://www.physics.umd.edu/enp/jjkelly/LEA
\bibitem{Cooper} E~.D.~Cooper, S.~Hama, B.~C.~Clark, and R.~L.~Mercer,
Phys. Rev. {\bf C47}, 297 (1993).
\bibitem{Fissum} K.~G.~Fissum {\it et al.}, Phys. Rev. {\bf C70}, 034606, 2004
\bibitem{Llew} C.~H.~Llewellyn Smith, Phys. Rep. {\bf 3C}, 1, 1972
\bibitem{Budd} H.~Budd, A.~Bodek, and J.~Arrington, Nucl. Phys. B. 
(Proc. Suppl) {\bf 139}, 90, (2005).
\bibitem{Flux} A.~A.~Aguilar-Arevalo {\it et al.}, (MiniBooNE 
Collaboration), Phys. Rev. {\bf D79}, 072002 (2009).
\bibitem{Sob2} J.~T.~Sobczyk, Eur. Phys. J. {\bf C72}, 1850 (2012).
\bibitem{Mann} W.~A.~Mann {\it et al.}, Phys. Rev. Lett. {\bf 31}, 844, (1973).
\bibitem{Baker} N.~J.~Baker {\it et al.}, Phys. Rev. {\bf D23}, 2499, (1981).
\bibitem{Pohl} M.~Pohl {\it et al.}, Lett. Nuovo Cim. {\bf 26}, 332, 1979.
\bibitem{Brunner} J.~Brunner {\it et al.}, Z. Phys. {\bf C45}, 551, 1990.
\bibitem{Kelly2} J.~J.~Kelly, Adv. Nucl. Phys. {\bf 23}, 75, 1996


\end{thebibliography}
\end{document}